\documentstyle[12pt,psfig]{article}

\makeatletter

\def\section{\@startsection {section}{1}{\z@}{24pt plus 2pt minus 2pt}
{12pt plus 2pt minus 2pt}{\large\bf}}

\def\subsection{\@startsection {subsection}{2}{\z@}{12pt plus 2pt minus 2pt}
{12pt plus 2pt minus 2pt}{\normalsize\bf}}
\makeatother

\begin{document}

\date{}


\title{\Large\bf \$7.0/Mflops Astrophysical $N$-Body Simulation \\
with Treecode on GRAPE-5}

\author{Atsushi Kawai\thanks{Department of General Systems Studies, 
College of Arts and Sciences, University of Tokyo, 
Tokyo 153, Email: kawai@grape.c.u-tokyo.ac.jp},  
Toshiyuki Fukushige\thanks{Department of General Systems Studies, 
College of Arts and Sciences, University of Tokyo, 
Tokyo 153, Email: fukushig@provence.c.u-tokyo.ac.jp}  
and Junichiro Makino\thanks{Department of Astronomy, 
School of Science, University of Tokyo, 
Tokyo 113, Email: makino@astron.s.u-tokyo.ac.jp}\\ 
University of Tokyo
}

\maketitle

\subsection*{\centering Abstract} 

As an entry for the 1999 Gordon Bell price/performance prize, we report
an astrophysical $N$-body simulation performed with a treecode on
GRAPE-5 (\underbar{Gra}vity \underbar{P}ip\underbar{e} 5) system, a
special-purpose computer for astrophysical $N$-body simulations.  The
GRAPE-5 system has 32 pipeline processors specialized for the
gravitational force calculation.  Other operations, such as tree
construction, tree traverse and time integration, are performed on a
general purpose workstation.  The total cost for the GRAPE-5 system is
40,900 dollars.  We performed a cosmological $N$-body simulation with
2.1 million particles, which sustained a performance of 5.92 Gflops
averaged over 8.37 hours.  The price per performance obtained is 7.0
dollars per Mflops. 

\section{Introduction}

Astrophysical $N$-body simulation is one of the most widely used
technique to investigate formation and evolution of astronomical
objects, such as galaxies, galaxy clusters and large scale structures of
the universe.  In such simulations, we calculate gravitational force on
each particle from all other particles, and integrate the orbit of each
particle according to Newton's equation of motion.  We investigate
structural and dynamical properties of the simulated object. 

The astrophysical $N$-body simulation has been one of grand challenge
problems in computational sciences.  In years 1992, 96, 97, and 98, the
Gordon Bell prizes were awarded to cosmological $N$-body simulations
\cite{gb92}\cite{gb96}\cite{gb97}\cite{gb98} and in 1995 the Gordon Bell
prize is awarded to $N$-body simulation of a black hole binary in a
galaxy \cite{gb95}.  The calculation cost of the astrophysical
$N$-body simulation rapidly increases for large $N$, because it is
proportional to $N^2$ if we use a straightforward approach.  The
gravity is a long-range attractive force.  A particle feel the forces
from all other particles, no matter how they are far away.  We cannot
use a cutoff technique which is widely used in MD simulation (e.g.
\cite{md}).  In order to reduce the calculation costs, various fast
algorithms have been developed.

Hierarchical tree algorithm \cite{bh} is one of such fast algorithms
which reduce the calculation cost from $O(N^2)$ to $O(NlogN)$.  In this
algorithm, particle are organized in the form of a tree, and each node
of the tree represents a group of particles.  The force from a distance
node is replaced by the force from its center of mass.  The Gordon Bell
prizes of years 1992, 97 and 98 were awarded to $N$-body simulations
with this tree algorithm \cite{gb92}\cite{gb97}\cite{gb98}, which were
performed on Intel Touchstone Delta, ASCI-Red, PC cluster, and an Alpha cluster. 

We report an astrophysical $N$-body simulation with the tree algorithm
on GRAPE-5 (GRAvity PipE) special-purpose computer.  GRAPE-5 has
dedicated pipelines specialized for the calculation of the gravitational
force.  It is connected to a host computer, which is general purpose
workstation, and operates as a hardware accelerator for the calculation
of the gravitational force.  Other operations, such as tree
construction, tree traverse and time integration, are performed on the
host computer.  It has been already demonstrated that the approach using
special-purpose machines successfully achieved very high performance in
scientific computations, by the Gordon Bell prize simulations of 1995
and 96 \cite{gb95}\cite{gb96}, which were performed on GRAPE-4\cite{g4},
and the last Gordon Bell prize simulation \cite{gb98b}, which was
performed on QCDSP\cite{qcdsp}. 
	
We performed a cosmological 2.1 million particles simulation using the
tree algorithm on GRAPE-5 connected to a COMPAQ AlphaServer DS10.  Sustained
performance is 5.92 Gflops and price/performance is {\bf \$7.0/Mflops}.  
In the rest of this paper, we describe on GRAPE-5 system
and the tree algorithm on GRAPE, and report the cost and performance. 

\section{GRAPE-5 system}

We briefly describe architecture of the GRAPE-5 system.  More detailed
descriptions of the GRAPE-5 system will be given elsewhere
\cite{g5}. GRAPE-5 is designed to run the tree code with very high
speed.  Figure 1 summarizes the configuration of the GRAPE-5 system
used for the simulation reported in this paper.  The GRAPE-5 system
consists of 2 processor boards, 2 host interface boards, and a host
computer.  The processor board performs the force calculation.  The
host interface board handles the communication between the processor
board and the host computer.  The host computer performs all other
operations.  We used COMPAQ AlphaServer DS10 with a 21264/466MHz Alpha
processor for the host computer. Figure 2 and figure 3 are photographs
of the GRAPE-5 system and GRAPE-5 processor board, respectively.

\begin{figure}
\centerline{\psfig{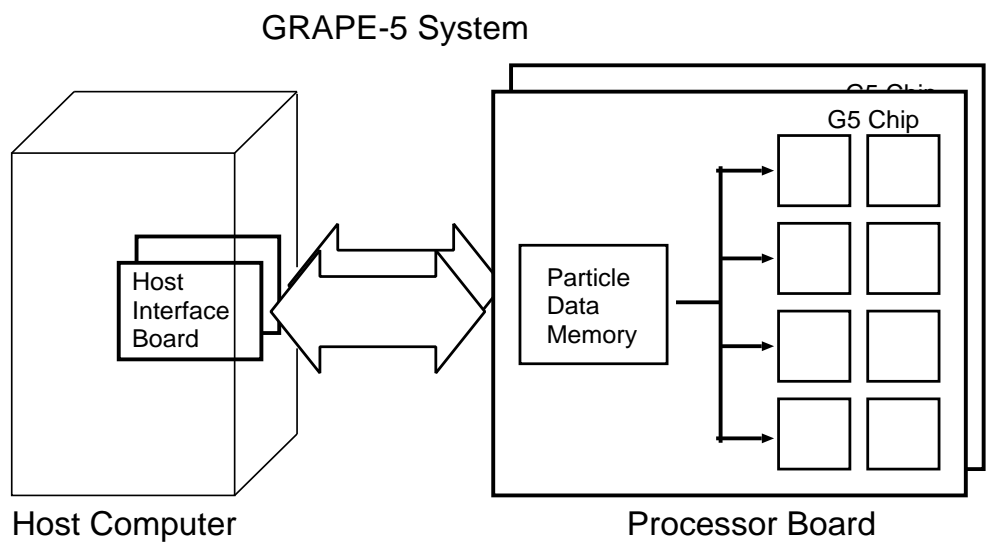}}
\caption{Block diagram of the GRAPE-5 system}
\end{figure}

\begin{figure}
\centerline{\psfig{file=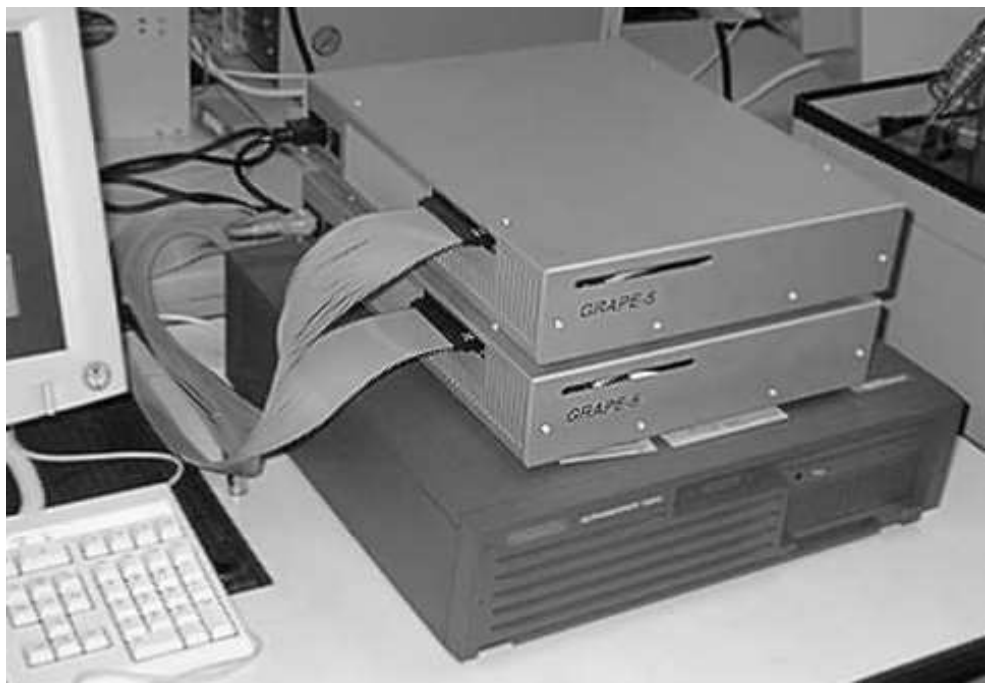,width=130mm}}
\caption{Photograph of the GRAPE-5 system}
\end{figure}

\begin{figure}
\centerline{\psfig{file=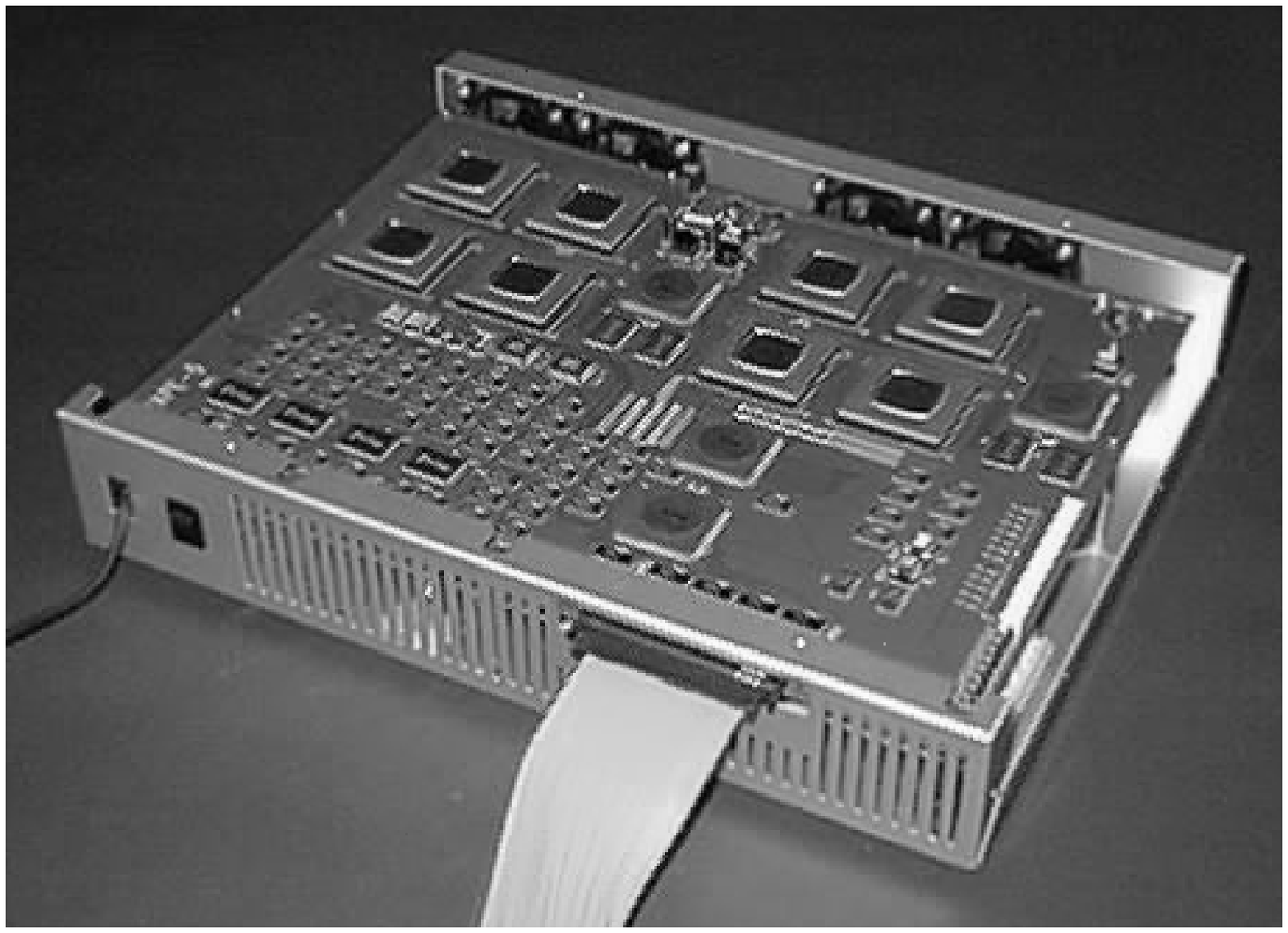,width=130mm}}
\caption{Photograph of the GRAPE-5 processor board}
\end{figure}

Each processor board consists of 8 processor chips (G5 chip) and a
particle data memory.  G5 chip is a custom LSI chip which calculates the
gravitational force.  Each G5 chip houses 2 pipelines specialized for
the force calculation.  The particle data memory stores the data of
particles which exert the force and supplies them to G5 chip.  G5 chip
operates at 90MHz and other part of the processor boards operate at
15MHz. 

G5 chip is designed for astrophysical $N$-body simulations with the tree
algorithm and calculates a pair-wise force with a relative error of about
0.3\%.  This might sound rather low, but detailed theoretical analysis
\cite{gerror} and numerical experiment \cite{hhm93} have shown that it
is more than enough.  The average error of the force in our simulation
is around 0.1\%, which is dominated by the approximation made in the
tree algorithm and not by the accuracy of the hardware.  The relative
accuracy was practically the same when we performed the same force
calculation using standard 64-bit floating point arithmetic.

The theoretical peak speed of the GRAPE-5 system is 109.44 Gflops.
Total number of pipeline processors is 32.  Each processor pipeline
operates 38 operations in a clock cycle, if we use the same counting
convention as used in \cite{gb97}\cite{gb98}.

\section{Tree algorithm}

Our code \cite{jmtree} is based on the Barnes's modified tree algorithm
\cite{b90}.  The implementation of the modified tree algorithm on GRAPE
were discussed in \cite{jm91}.  Using this algorithm, the calculation
cost on the host computer is greatly reduced from that of the original
algorithm and the forces exerted on multiple particles can be calculated
in parallel.  In the original algorithm, the interaction list is created
for each particle.  In the modified tree algorithm, neighboring
particles are grouped and one interaction list is shared among the
particles in the same group.  Forces from particles in the same group is
directly calculated. 

The modified tree algorithm reduces the calculation cost of the host
computer by roughly a factor of $n_g$, where $n_g$ is the average number
of particles in a group.  On the other hand, the amount of work on
GRAPE-5 increases as we increase $n_g$, since the interaction list
becomes longer.  There is, therefore, an optimal $n_g$ at which the
total computing time is minimum.  The optimal $n_g$ strongly depends on
the ratio of the speed of the host computer and GRAPE.  For the present
configuration, the optimal $n_g$ is around 2000. 

Note that our modified tree algorithm performs larger number of
operations than the tree algorithm on a general purpose computer.
When we will estimate the performance in section 5, we will make
correction. Note also that the our modified tree algorithm is more
accurate than the original tree algorithm for the same accuracy
parameter, as shown in \cite{b90}\cite{km99}. 
 
\section{Cost}

The total cost of the GRAPE-5 system is 4.7 M JYE.  The GRAPE-5 board
is available from a Japanese commercial company for the price of 1.65
M JYE per board.  Remaining 1.4 M JYE was spent for the host computer,
COMPAQ AlphaServer DS10, including 512 MByte main memory and C++
compiler.  The total cost, with the present exchange rate of 1 dollar
= 115 JYE, is about 40,900 dollars.

\section{Simulation}

We report the performance statistics for the astrophysical $N$-body
simulations with the tree algorithm on GRAPE-5.  The performance numbers
are based on the wall-clock time obtained from UNIX system timer on the
host computer (COMPAQ AlphaServer DS10). 

We performed a cosmological $N$-body simulation of a sphere of radius
50Mpc (mega parsec) with 2,159,038 particles for 999 timesteps. We
assigned the initial position and velocities to particles in a
spherical region selected from a discrete realization of density
contrast field based on a standard cold dark matter scenario using
COSMICS package
\cite{cosmic}.  A particle represents $1.7\times 10^{10}$ solar masses. 
We performed the simulation from $z=24$, where $z$ is redshift, to the
present time. Figure 4 shows a snapshot of the simulation.

\begin{figure}
\centerline{\psfig{file=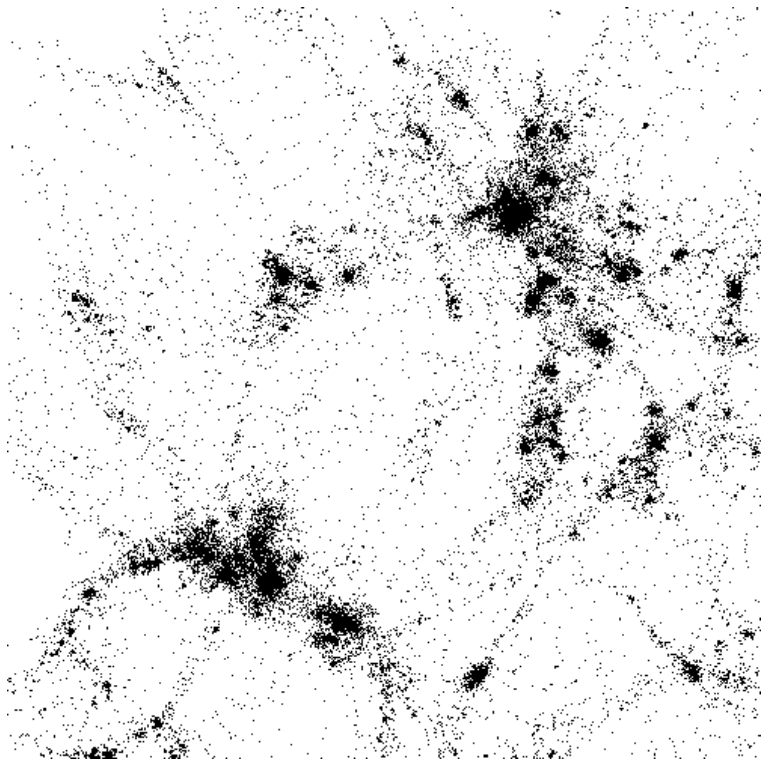,width=100mm}}
\caption{A snapshot of the simulation
at $z=0$ (present time).  Particles in a 45Mpc $\times$ 45Mpc $\times$
2.5Mpc box are plotted.}
\end{figure}

The total number of the particle-particle interactions is $2.90 \times
10^{13}$.  This implies that the average length of the interaction list
is 13,431.  The whole simulation took 30,141 seconds (8.37 hours)
including I/O, resulting in the average computing speed of 36.4 Gflops. 
Here we use the operation count of 38 per interaction. 

However, as we described in section 3, our modified tree algorithm
performs larger number of operations than the tree algorithm on a
general purpose computer.  In order to make correction, we estimated the
operation count of the original tree algorithm for the same simulation,
using five snapshot files and the same accuracy parameter.  The
estimated number of the interaction is $4.69\times 10^{12}$.  The
effective sustained speed is {\bf 5.92 Gflops} and the price/performance
is {\bf \$7.0/Mflops}.


\begin{thebibliography}{100}

\bibitem{gb92}
M. S. Warren and J. K. Salmon, in {\it Proceedings
of Supercomputing '92}, IEEE Computer Society Press, Los Alamitos,
1992, p 570-576.

\bibitem{gb96}
T. Fukushige  and J. Makino, in {\it Proceedings
of Supercomputing '96}, IEEE Computer Society Press, Los Alamitos,
1996. 

\bibitem{gb97} 
M. S. Warren, J. K. Salmon, D. J. Becker, M. P. Goda, T. Sterling, 
and  G. S. Winckelmans, in {\it Proceedings of Supercomputing
'97}, IEEE Computer Society Press, Los Alamitos, 1997. 

\bibitem{gb98}
M. S. Warren, T. C. Germann, P. S. Lamdahl,  D. M. Beazley, J. K. Salmon, 
in {\it Proceedings of Supercomputing '98}, IEEE Computer Society Press, Los Alamitos,
1998. 

\bibitem{gb95}
J.  Makino and M.  Taiji, in {\it Proceedings of Supercomputing '95},
IEEE Computer Society Press, Los Alamitos, 1995. 

\bibitem{md}
Y. Duan  and P. A. Kollman, {\it Science}, Vol. 282, p. 740, 1998.

\bibitem{bh}
J. Barnes, and P. Hut, {\it  Nature},  Vol. 324, p. 446,  1986.

\bibitem{g4}
J. Makino, M. Taiji, T. Ebisuzaki, and D. Sugimoto, 
{\it Astrophysical Journal}, Vol. 480, p. 432, 1997.    

\bibitem{gb98b}
D. Chen, et al.  in {\it Proceedings of Supercomputing '98}, IEEE
Computer Society Press, Los Alamitos, 1998. 

\bibitem{qcdsp}
D. Chen, et al. in {\it Proceedings
of Supercomputing '97}, IEEE Computer Society Press, Los Alamitos,
1997. 

\bibitem{g5} 
A.  Kawai, T.  Fukushige, J.  Makino, and M.  Taiji, to be submitted to
{\it Publ.  Astron.  Soc.  Japan}, 1999. 

\bibitem{gerror}
J. Makino, T. Ito, and T. Ebisuzaki, 
{\it Publ. Astron. Soc. Japan}, Vol. 42 p. 717,  1990.

\bibitem{hhm93}
L. Hernquist, P. Hut, and J. Makino, 
{\it Astrophysical Journal}, Vol. 402, p. L85, 1993.

\bibitem{jmtree}
J. Makino, {\tt http://grape.c.u-tokyo.ac.jp/\verb+~+makino/softwares/C++tree}
{\tt /index.html}

\bibitem{b90}
J. Barnes, {\it J. Comp. Phys.},  Vol 87, p. 161,  1990.

\bibitem{jm91}
J. Makino, {\it Publ. Astron. Soc. Japan}, Vol. 43 p. 621,  1991.

\bibitem{km99} 
A.  Kawai and J.  Makino, in {\it Proceedings of the Ninth SIAM
Conference on Parallel Processing for Scientific Computing}, SIAM, 1999. 

\bibitem{cosmic}
E. Bertschinger, {\tt http://arcturus.mit.edu:80/cosmics/}

\end{thebibliography}
\end{document}